\documentclass[10pt,a4paper,twocolumn,english,conference]{IEEEtran}
\usepackage[T1]{fontenc}
\usepackage[latin9]{inputenc}
\usepackage{float}
\usepackage{amsmath}
\usepackage{amssymb}
\usepackage{stackrel}
\usepackage{graphicx}
\usepackage{esint}

\makeatletter

\pdfpageheight\paperheight
\pdfpagewidth\paperwidth

\floatstyle{ruled}
\newfloat{algorithm}{tbp}{loa}
\providecommand{\algorithmname}{Algorithm}
\floatname{algorithm}{\protect\algorithmname}

\usepackage{cite}\usepackage{algorithm}\usepackage{algorithmicx}\usepackage{algpseudocode}\usepackage{babel}

\allowdisplaybreaks \allowdisplaybreaks[4]

\usepackage{babel}

\makeatother

\usepackage{babel}
\begin{document}

\title{Concise Derivation of Complex Bayesian Approximate Message Passing
via Expectation Propagation }

\author{\IEEEauthorblockN{Xiangming Meng\IEEEauthorrefmark{1}\IEEEauthorrefmark{3},
Sheng Wu\IEEEauthorrefmark{2}, Linling Kuang\IEEEauthorrefmark{2},
Jianhua Lu\IEEEauthorrefmark{1}\IEEEauthorrefmark{3} }\IEEEauthorblockA{\IEEEauthorrefmark{1}Department
of Electronic Engineering, Tsinghua University, Beijing, China }\IEEEauthorblockA{\IEEEauthorrefmark{2}Tsinghua
Space Center, Tsinghua University, Beijing, China} \IEEEauthorblockA{\IEEEauthorrefmark{3}Tsinghua
National Laboratory for Information Science and Technology, Beijing,
China}\IEEEauthorblockA{Email: mengxm11@mails.tsinghua.edu.cn,
\{thuraya, kll, lhh-dee\}@mail.tsinghua.edu.cn}}
\maketitle
\begin{abstract}
In this paper, we address the problem of recovering complex-valued
signals from a set of complex-valued linear measurements. \textit{Approximate
message passing} (AMP) is one state-of-the-art algorithm to recover
real-valued sparse signals. However, the extension of AMP to complex-valued
case is nontrivial and no detailed and rigorous derivation has been
explicitly presented. To fill this gap, we extend AMP to complex Bayesian
approximate message passing (CB-AMP) using expectation propagation
(EP). This novel perspective leads to a concise derivation of CB-AMP
without sophisticated transformations between the complex domain and
the real domain. In addition, we have derived state evolution equations
to predict the reconstruction performance of CB-AMP. Simulation results
are presented to demonstrate the efficiency of CB-AMP and state evolution. \end{abstract}


\section{Introduction}

Compressed sensing (CS) aims to undersample high-dimensional signals
yet accurately reconstruct them by exploiting their structure\cite{Donoho-CompressiveSensing,Candes-Introduction-to-CS}.
To this end, a plethora of methods have been proposed in the past
years \cite{eldar2012compressed}. Among others, approximate message
passing (AMP) \cite{donoho2009message} proposed by Donoho \textit{et
al.} is one state-of-the-art algorithm to recover sparse signals.
As an efficient application of belief propagation\cite{Pearl1988,Kschischang2001},
AMP has found various applications in solving linear inverse problems.
Moreover, AMP has been extended to Bayesian AMP (B-AMP) \cite{donoho2010message,krzakala2012probabilistic}
and general linear mixing problems\cite{rangan2011generalized,schniter2011message,Rangan-AMP-Learning}.
However, most of the existing works focus on the case of real-valued
signals and measurements, while in many applications, e.g., communication\cite{schniter2011message},
magnetic resonance imaging\cite{vlaardingerbroek2013magnetic}, and
radar imaging\cite{baraniuk2007compressive}, etc., it is more convenient
to represent signals in the complex-domain\cite{yang2012phase}. Though
it can be transformed and processed in the real domain, it is beneficial
to deal with complex-valued signals in a straightforward way since
their real and imaginary components are often either both zero or
both non-zero simultaneously\cite{maleki2013asymptotic,yang2012phase}.

The extension of AMP to deal with complex-valued signals with complex-valued
measurements has already been considered in \cite{schniter2011message,som2012compressive,vila2013expectation,maleki2013asymptotic,barbier2015approximate}.
In \cite{maleki2013asymptotic}, the authors proposed one kind of
complex approximate message passing (CAMP) algorithm. However, the
extension of AMP to CAMP is sophisticated. A more compact form of
CAMP is proposed in \cite{schniter2011message,som2012compressive,vila2013expectation,barbier2015approximate}.
To the best of our knowledge, although such extensions have been considered,
no detailed and rigorous derivation has been explicitly presented.
In \cite{meng2015expectation}, we derived the original AMP algorithm
from the \textit{expectation propagation} (EP)\cite{minka2001family,minka2005divergence}
perspective, which unveils the intrinsic connection between AMP and
EP. Nevertheless, it only deals with real-valued sparse signals with
Laplace prior, which limits its use in more general problems. In this
paper we further extend it to complex Bayesian AMP (CB-AMP), i.e.,
complex-valued signal reconstruction with general known prior distribution.
This novel perspective leads to a concise and natural extension from
AMP to CB-AMP, without sophisticated transformations between the complex
domain and the real domain. In addition, we have also derived state
evolution equations to predict the reconstruction performance of CB-AMP.
The superiority of CB-AMP is demonstrated via simalation results,
which are consistent with the prediction results of state evolution
equations.

\section{Derivation of CB-AMP via EP }

\subsection{System Model}

Consider a complex-valued linear system of the form
\begin{equation}
\mathbf{y}=\mathbf{Ax}+\mathbf{w},\label{eq:linear_system}
\end{equation}
where $\mathbf{x}\in\mathbb{C}^{N}$ is the unknown complex signal,
$\mathbf{A}\in\mathbb{C}^{M\times N}$ is the measurement matrix,
$\mathbf{w}\in\mathbb{C}^{M}$ is the additive complex Gaussian noise
with zero mean and covariance matrix $\sigma^{2}\mathbf{I}_{M}$,
where $\mathbf{I}_{M}$ is the identity matrix of size $M$. The complex
Gaussian distribution of $\mathbf{w}$ is denoted by $\mathcal{CN}\bigl(\mathbf{w};0,\sigma^{2}\mathbf{I}_{M}\bigr)$.
The prior distribution of signal $\mathbf{x}$ is supposed to be known
and has a separable form
\begin{equation}
p_{0}\bigl(\mathbf{x}\bigr)=\stackrel[i=1]{N}{\prod}p_{0}\bigl(x_{i}\bigr).\label{eq:prior_dist}
\end{equation}

The goal is to estimate $\mathbf{x}$ from the noisy observations
$\mathbf{y}$ given $\mathbf{A}$ and the statistical information
of $\mathbf{x}$ and $\mathbf{w}$ using the minimum mean square error
(MMSE) criterion. It is well known that the MMSE estimate of $x_{i}$
is the posterior mean, i.e., $\hat{x}_{i}=\int x_{i}p\bigl(x_{i}|\mathbf{y}\bigr)dx_{i},$
where $p\bigl(x_{i}|\mathbf{y}\bigr)$ is the marginal distribution
of the joint posterior distribution 
\begin{align}
p\bigl(\mathbf{x}|\mathbf{y}\bigr) & =\frac{p\bigl(\mathbf{y}|\mathbf{x}\bigr)p_{0}\bigl(\mathbf{x}\bigr)}{p\bigl(\mathbf{y}\bigr)}\nonumber \\
 & \propto\stackrel[a=1]{M}{\prod}p\bigl(y_{a}|\mathbf{x}\bigr)\stackrel[i=1]{N}{\prod}p_{0}\bigl(x_{i}\bigr),\label{eq:poster_prob}
\end{align}
where $\propto$ denotes identity between two distributions up to
a normalization constant. Under the statistical assumption of measurement
noise $\mathbf{\mathbf{w}}$, the conditional distribution of the
$a$-th element of $\mathbf{y}$, $y_{a}$, given $\mathbf{x}$ can
be explicitly represented as
\begin{equation}
p\bigl(y_{a}|\mathbf{x}\bigr)=\frac{1}{\pi\sigma^{2}}\exp\Bigl(-\frac{1}{\sigma^{2}}\bigl|y_{a}-\underset{i}{\sum}A_{ai}x_{i}\bigr|^{2}\Bigr).\label{eq:likelihood_prob}
\end{equation}

Message passing algorithms \cite{Pearl1988,Kschischang2001,wainwright2008graphical}
provide a family of efficient methods to (approximately) compute the
marginals. The basic paradigm is well illustrated via factor graph
\cite{Kschischang2001} which represents the statistical dependencies
between random variables. The factorization in (\ref{eq:poster_prob})
can be encoded in a factor graph $\mathcal{G}\!=\!\!\bigl(\mathcal{V},\mathcal{F},\mathcal{E}\bigr)$,
where $\mathcal{V}=\left\{ i\right\} $ is the set of variable nodes,
$\mathcal{F}=\left\{ a\right\} $ is the set of factor nodes and $\mathcal{E}$
denotes the set of edges. In the sequel, we assume that the measurement
matrix $\mathbf{A}$ is a homogenous matrix whose elements admit i.i.d.
distribution with mean zero and variance $\gamma$.

\subsection{Approximate inference using EP}

Given the factor graph representation, the marginals can be computed
distributively via local message passing\cite{Kschischang2001,minka2005divergence,minka2001family}.
The projection of a particular distribution $p$ into a distribution
set $\Phi$ is defined as \cite{minka2005divergence} 
\begin{equation}
\mathrm{Proj_{\Phi}}\left[p\right]=\mathrm{arg}\,\mathrm{\min_{\mathit{q}\in\Phi}}\, D\left(p||q\right),\label{eq:Projection rule-1}
\end{equation}
where $D\left(p||q\right)$ denotes the Kullback-Leibler divergence. 

Denote by $m_{i\rightarrow a}^{t}\bigl(x_{i}\bigr)$ and $m_{a\rightarrow i}^{t}\bigl(x_{i}\bigr)$
the message from variable node $i$ to factor node $a$ in the $t$th
iteration and the message in the opposite direction, respectively.
Then, the message passing update rules of EP read\cite{minka2001family,minka2005divergence}
\begin{alignat}{1}
m_{i\rightarrow a}^{t+1}\bigl(x_{i}\bigr) & \propto\frac{\textrm{Proj}{}_{\Phi}\bigl[p_{0}\bigl(x_{i}\bigr)\prod_{b}m_{b\rightarrow i}^{t}\bigl(x_{i}\bigr)\bigr]}{m_{a\rightarrow i}^{t}\bigl(x_{i}\bigr)},\label{eq:itoa}\\
m_{a\rightarrow i}^{t}\bigl(x_{i}\bigr) & \propto\frac{1}{m_{i\rightarrow a}^{t}\bigl(x_{i}\bigr)}\textrm{Proj}{}_{\Phi}\Bigl[m_{i\rightarrow a}^{t}\left(x_{i}\right)\nonumber \\
 & \hspace{1em}\times\int\prod_{j\neq i}m_{j\rightarrow a}^{t}\bigl(x_{j}\bigr)p\bigl(y_{a}|\mathbf{x}\bigr)\Bigr].\label{eq:atoi}
\end{alignat}

After projection, each message $m_{j\rightarrow a}^{t}\bigl(x_{j}\bigr)$
from variable node $j$ to factor node $a$ is approximated as complex
Gaussian density function $\mathcal{CN}\bigl(x_{j};\hat{x}_{j\rightarrow a}^{t},\nu_{j\rightarrow a}^{t}\bigr)$,
thus, under the product measure $\prod_{j\neq i}m_{j\rightarrow a}^{t}\bigl(x_{j}\bigr)$,
the random variables $x_{j},\: j\!\!\neq\!\! i$ are independent complex
Gaussian random variables. Define $Z_{ai}=\underset{j\neq i}{\sum}A_{aj}x_{j}$,
so that $Z_{ai}$ is a complex Gaussian random variable with mean
and variance, respectively, 
\begin{alignat}{1}
Z_{a\rightarrow i}^{t} & =\sum_{j\neq i}A_{aj}\hat{x}_{j\rightarrow a}^{t},\label{eq:Z_a2i}\\
V_{a\rightarrow i}^{t} & =\sum_{j\neq i}|A_{aj}|^{2}\nu_{j\rightarrow a}^{t}.\label{eq:V_a2i}
\end{alignat}

Then, we obtain 
\begin{align}
 & \int\prod_{j\neq i}m_{j\rightarrow a}^{t}\bigl(x_{j}\bigr)p\bigl(y_{a}|\mathbf{x}\bigr)\nonumber \\
 & \propto\mathcal{CN}\bigl(x_{i};\frac{y_{a}-Z_{a\rightarrow i}^{t}}{A_{ai}},\frac{\sigma^{2}+V_{a\rightarrow i}^{t}}{|A_{ai}|^{2}}\bigr),
\end{align}
which implies that $x_{i}$ also admits complex Gaussian distribution.
In this case, the projection operation in (\ref{eq:atoi}) reduces
to identity operation, so that 
\begin{align}
m_{a\rightarrow i}^{t}\bigl(x_{i}\bigr) & \propto\mathcal{CN}\bigl(x_{i};\frac{y_{a}-Z_{a\rightarrow i}^{t}}{A_{ai}},\frac{\sigma^{2}+V_{a\rightarrow i}^{t}}{|A_{ai}|^{2}}\bigr).\label{eq:atoi_message}
\end{align}

Next we evaluate the message $m_{i\rightarrow a}^{t+1}\bigl(x_{i}\bigr)$.
The marginal posterior density estimate of $x_{i}$, i.e., the marginal
belief estimate $m_{i}^{t+1}\bigl(x_{i}\bigr)$, is defined as
\begin{equation}
m_{i}^{t+1}\bigl(x_{i}\bigr)=\textrm{Proj}{}_{\Phi}\Bigl[p_{0}\bigl(x_{i}\bigr)\prod_{a}m_{a\rightarrow i}^{t}\left(x_{i}\right)\Bigr].\label{eq:a posteriori pro density}
\end{equation}

According to the product rule of Gaussian functions\cite{bishop2006pattern},we
have 
\begin{alignat}{1}
\prod_{a}m_{a\rightarrow i}^{t}\left(x_{i}\right) & \propto\mathcal{CN}\bigl(x_{i};R_{i}^{t},\Sigma_{i}^{t}\bigr),\label{eq:product of gaus}
\end{alignat}
where
\begin{alignat}{1}
\Sigma_{i}^{t} & =\Bigl[\sum_{a}\frac{|A_{ai}|^{2}}{\sigma^{2}+V_{a\rightarrow i}^{t}}\Bigr]^{-1},\label{eq:invision}\\
R_{i}^{t} & =\Sigma_{i}^{t}\sum_{a}\frac{A_{ai}^{*}\bigl(y_{a}-Z_{a\rightarrow i}^{t}\bigr)}{\sigma^{2}+V_{a\rightarrow i}^{t}},\label{eq:invmean}
\end{alignat}
and $\left(\cdot\right)^{*}$ denotes conjugate operation.

For notational brevity, we introduce a family of density functions
\begin{equation}
p\left(x;R,\Sigma\right)\equiv\frac{p_{0}\bigl(x\bigr)}{z\left(R,\Sigma\right)}\exp\Bigl[-\frac{|x-R|^{2}}{\Sigma}\Bigr],\label{eq:distribution family f}
\end{equation}
where $z\left(R,\Sigma\right)=\int p_{0}\bigl(x\bigr)\exp\bigl(-|x-R|^{2}/\Sigma\bigr)dx$
is the normalization constant. As in \cite{krzakala2012probabilistic,barbier2015approximate},
the corresponding mean and variance are denoted as
\begin{align}
f_{a}\left(R,\Sigma\right) & =\int xp\left(x;R,\Sigma\right)dx,\label{eq:mean_def}\\
f_{c}\left(R,\Sigma\right) & =\int|x-f_{a}\left(R,\Sigma\right)|^{2}p\left(x;R,\Sigma\right)dx.\label{eq:var_def}
\end{align}

Combining $\left(\ref{eq:product of gaus}\right)$, $\left(\ref{eq:distribution family f}\right)$,
$\left(\ref{eq:mean_def}\right)$ and (\ref{eq:var_def}), we obtain
the tentative approximation of the posterior mean and variance of
$x_{i}$ in the $(t+1)$th iteration, which are denoted by $f_{a}\left(R_{i}^{t},\Sigma_{i}^{t}\right)$
and $f_{c}\left(R_{i}^{t},\Sigma_{i}^{t}\right)$, respectively. Then,
using projection operation $\left(\ref{eq:a posteriori pro density}\right)$
and moment matching, we project the posterior belief to the complex
Gaussian distribution set, yielding
\begin{equation}
m_{i}^{t+1}\bigl(x_{i}\bigr)\propto\mathcal{CN}\bigl(x_{i};\hat{x}_{i}^{t+1},\hat{\nu}_{i}^{t+1}\bigr),\label{eq:post_prob}
\end{equation}
where 
\begin{align*}
\hat{x}_{i}^{t+1} & =f_{a}\left(R_{i}^{t},\Sigma_{i}^{t}\right),\\
\hat{\nu}_{i}^{t+1} & =f_{c}\left(R_{i}^{t},\Sigma_{i}^{t}\right).
\end{align*}

According to $\left(\ref{eq:itoa}\right)$ and $\left(\ref{eq:a posteriori pro density}\right)$,
the message from variable node $i$ to factor node $a$ is evaluated
by 
\begin{align}
m_{i\rightarrow a}^{t+1}\left(x_{i}\right) & \propto\mathcal{CN}\bigl(x_{i};\hat{x}_{i\rightarrow a}^{t+1},\nu_{i\rightarrow a}^{t+1}\bigr),\label{eq:mesageitoa}
\end{align}
where 
\begin{alignat}{1}
\frac{1}{\nu_{i\rightarrow a}^{t+1}} & =\frac{1}{\nu_{i}^{t+1}}-\frac{|A_{ai}|^{2}}{\sigma^{2}+V_{a\rightarrow i}^{t}},\label{eq:var_i2a}\\
\hat{x}_{i\rightarrow a}^{t+1} & =\nu_{i\rightarrow a}^{t+1}\Bigl(\frac{\hat{x}_{i}^{t+1}}{\nu_{i}^{t+1}}-\frac{A_{ai}^{*}\bigl(y_{a}-Z_{a\rightarrow i}^{t}\bigr)}{\sigma^{2}+V_{a\rightarrow i}^{t}}\Bigr).\label{eq:mean_i2a}
\end{alignat}

Now we have closed the message computation. However, about $\mathcal{O}\left(MN\right)$
messages need to be computed. In the sequel, we further reduce the
number of messages per iteration to $\mathcal{O}\left(M+N\right)$
by neglecting the high order terms in large system limit.

\subsection{Reducing the number of messages}

Define 
\begin{alignat}{1}
Z_{a}^{t} & =\sum_{i}A_{ai}\hat{x}_{i\rightarrow a}^{t},\label{eq:Za_def}\\
V_{a}^{t} & =\sum_{i}|A_{ai}|^{2}\nu_{i\rightarrow a}^{t},\label{eq:Va_def}
\end{alignat}

Then, it can be easily seen that (\ref{eq:Z_a2i}) and (\ref{eq:V_a2i})
can be rewritten as
\begin{alignat}{1}
Z_{a\rightarrow i}^{t} & =Z_{a}^{t}-A_{ai}\hat{x}_{i\rightarrow a}^{t},\label{eq:Z_a2i-1}\\
V_{a\rightarrow i}^{t} & =V_{a}^{t}-|A_{ai}|^{2}\nu_{i\rightarrow a}^{t}.\label{eq:V_a2i-1}
\end{alignat}

Neglecting the high order term $|A_{ai}|^{2}\nu_{i\rightarrow a}^{t}$
in (\ref{eq:V_a2i-1}), we have
\begin{equation}
V_{a\rightarrow i}^{t}\approx V_{a}^{t},\label{eq:Va2i_app}
\end{equation}
which is independent of $i$.

The simplification of $Z_{a\rightarrow i}^{t}$ is not that trivial
since we should be careful to keep the Onsager reaction term in approximating
$Z_{a\rightarrow i}^{t}$. From (\ref{eq:var_i2a}), neglecting the
high order term $|A_{ai}|^{2}/(\sigma^{2}+V_{a\rightarrow i}^{t})$,
we obtain
\begin{equation}
\nu_{i\rightarrow a}^{t+1}\approx\nu_{i}^{t+1},\label{eq:var_approximate}
\end{equation}
so that 
\begin{equation}
V_{a}^{t}\approx\sum_{i}|A_{ai}|^{2}\nu_{i}^{t}.\label{eq:Va_app}
\end{equation}

Substituting $\left(\ref{eq:var_approximate}\right)$ and (\ref{eq:Va2i_app})
into $\left(\ref{eq:mean_i2a}\right)$, we have
\begin{equation}
\hat{x}_{i\rightarrow a}^{t+1}\approx\hat{x}_{i}^{t+1}-\nu_{i}^{t+1}\frac{A_{ai}^{*}\bigl(y_{a}-Z_{a\rightarrow i}^{t}\bigr)}{\sigma^{2}+V_{a}^{t}}.\label{eq:x_i2a_app}
\end{equation}

Combining (\ref{eq:Z_a2i-1}), (\ref{eq:Va2i_app}), and (\ref{eq:x_i2a_app}),
we have
\begin{align}
Z_{a\rightarrow i}^{t} & =Z_{a}^{t}-A_{ai}\hat{x}_{i}^{t}+\nu_{i}^{t}\frac{|A_{ai}|^{2}\bigl(y_{a}-Z_{a\rightarrow i}^{t-1}\bigr)}{\sigma^{2}+V_{a}^{t-1}},\label{eq:Z_a2i_app2}
\end{align}
which leads to a further approximation 
\begin{align}
\hat{x}_{i\rightarrow a}^{t+1} & \approx\hat{x}_{i}^{t+1}-\frac{\nu_{i}^{t+1}A_{ai}^{*}}{\sigma^{2}+V_{a}^{t}}\Bigl[y_{a}-\nonumber \\
 & \hspace{1em}Z_{a}^{t}+A_{ai}\hat{x}_{i}^{t}-\nu_{i}^{t}\frac{|A_{ai}|^{2}\bigl(y_{a}-Z_{a\rightarrow i}^{t-1}\bigr)}{\sigma^{2}+V_{a}^{t-1}}\Bigr]\nonumber \\
 & \approx\hat{x}_{i}^{t+1}-\frac{\bigl(y_{a}-Z_{a}^{t}\bigr)\nu_{i}^{t+1}A_{ai}^{*}}{\sigma^{2}+V_{a}^{t}},\label{eq:x_i2a_app2}
\end{align}
where the last step is approximated by neglecting the high order terms.
Then, $Z_{a}^{t}$ defined in (\ref{eq:Za_def}) can be approximated
as
\begin{align}
Z_{a}^{t} & \approx\sum_{i}A_{ai}\hat{x}_{i}^{t}-\frac{\bigl(y_{a}-Z_{a}^{t-1}\bigr)}{\sigma^{2}+V_{a}^{t-1}}V_{a}^{t}.\label{eq:Update_Za}
\end{align}

Substituting (\ref{eq:Va2i_app}) into (\ref{eq:invision}) leads
to 
\begin{alignat}{1}
\Sigma_{i}^{t} & \approx\Bigl[\sum_{a}\frac{|A_{ai}|^{2}}{\sigma^{2}+V_{a}^{t}}\Bigr]^{-1}.\label{eq:Vi_app}
\end{alignat}

Substituting (\ref{eq:Va2i_app}) and (\ref{eq:Z_a2i_app2}) into
(\ref{eq:invmean}), we have
\begin{align}
R_{i}^{t} & \approx\Sigma_{i}^{t}\sum_{a}\frac{A_{ai}^{*}\bigl(y_{a}-Z_{a}^{t}\bigr)}{\sigma^{2}+V_{a}^{t}}+\Sigma_{i}^{t}\sum_{a}\frac{|A_{ai}|^{2}}{\sigma^{2}+V_{a}^{t}}\hat{x}_{i}^{t}\nonumber \\
 & \hspace{1em}-\Sigma_{i}^{t}\sum_{a}\frac{A_{ai}^{*}}{\sigma^{2}+V_{a}^{t}}\nu_{i}^{t}\frac{|A_{ai}|^{2}\bigl(y_{a}-Z_{a\rightarrow i}^{t-1}\bigr)}{\sigma^{2}+V_{a}^{t-1}}\nonumber \\
 & \approx\hat{x}_{i}^{t}+\Sigma_{i}^{t}\sum_{a}\frac{A_{ai}^{*}\bigl(y_{a}-Z_{a}^{t}\bigr)}{\sigma^{2}+V_{a}^{t}},\label{eq:Update_Ri}
\end{align}
where in the last step we have neglected high order term and used
the relationship (\ref{eq:Vi_app}).

At this step, we finally obtain the complex Bayesian approximate message
passing (CB-AMP) as shown in algorithm\ref{CAMP}, which is the same
as that in \cite{barbier2015approximate} and \cite{schniter2011message}
( note that some notational modification is needed to match \cite{schniter2011message}).

\begin{algorithm}
\protect\caption{CB-AMP}

\begin{raggedright}
1) Initialization: $t=1,\hat{x}_{i}^{1}=\int x_{i}p_{0}(x_{i})dx,\nu_{i}^{1}=\int|x_{i}-\hat{x}_{i}^{1}|p_{0}(x_{i})dx,i=1,\ldots,N$,$V_{a}^{0}=1,Z_{a}^{0}=y_{a},a=1,\ldots,M.$
\par\end{raggedright}

2) Factor node update: For $a=1,\ldots,M$
\begin{align*}
V_{a}^{t} & =\sum_{i}|A_{ai}|^{2}\nu_{i}^{t},\\
Z_{a}^{t} & =\sum_{i}A_{ai}\hat{x}_{i}^{t}-\frac{V_{a}^{t}}{\sigma^{2}+V_{a}^{t-1}}\bigl(y_{a}-Z_{a}^{t-1}\bigr).
\end{align*}

3) Variable node update: For $i=1,\ldots N$
\begin{alignat*}{1}
\Sigma_{i}^{t} & =\Bigl[\sum_{a}\frac{|A_{ai}|^{2}}{\sigma^{2}+V_{a}^{t}}\Bigr]^{-1},\\
R_{i}^{t} & =\hat{x}_{i}^{t}+\Sigma_{i}^{t}\sum_{a}\frac{A_{ai}^{*}\bigl(y_{a}-Z_{a}^{t}\bigr)}{\sigma^{2}+V_{a}^{t}},\\
\hat{x}_{i}^{t+1} & =f_{a}\left(R_{i}^{t},\Sigma_{i}^{t}\right),\\
\hat{\nu}_{i}^{t+1} & =f_{c}\left(R_{i}^{t},\Sigma_{i}^{t}\right).
\end{alignat*}

4) Set $t\leftarrow t+1$ and proceed to step 2) until a predefined
number of iterations or other termination conditions are satisfied.
\label{CAMP}
\end{algorithm}

\section{State Evolution Analysis }

We are interested in the mean square error (MSE) to characterize the
reconstruction performance, which is defined as
\begin{equation}
MSE=\frac{1}{N}\stackrel[i=1]{N}{\sum}\bigl|\hat{x}_{i}-x_{i}\bigr|^{2}.\label{eq:MSE_def}
\end{equation}

As is shown in \cite{krzakala2012probabilistic}, the state evolution
(or cavity method) uses a statistical analysis of the messages at
iteration $t$, in the large system limit, to derive their distributions
at iteration $t+1$. Define
\begin{alignat}{1}
V^{t} & =\frac{1}{N}\stackrel[i=1]{N}{\sum}\nu_{i}^{t},\qquad E^{t}=\frac{1}{N}\stackrel[i=1]{N}{\sum}\bigl|\hat{x}_{i}^{t}-x_{i}\bigr|^{2}.\label{eq:aver-var}
\end{alignat}

We first focus on the calculation of $R_{i}^{t}$. Substituting (\ref{eq:Z_a2i}),
(\ref{eq:V_a2i}), (\ref{eq:invision}) as well as the system model
(\ref{eq:linear_system}) into the definition (\ref{eq:invmean}),
we have
\begin{align}
R_{i}^{t} & =\frac{\sum_{a}\frac{|A_{ai}|^{2}x_{i}+A_{ai}^{*}w_{a}+A_{ai}^{*}\sum_{j\neq i}A_{aj}\bigl(x_{j}-\hat{x}_{j\rightarrow a}^{t}\bigr)}{\sigma^{2}+\sum_{j\neq i}|A_{aj}|^{2}\nu_{j\rightarrow a}^{t}}}{\sum_{a}\frac{|A_{ai}|^{2}}{\sigma^{2}+\sum_{j\neq i}|A_{aj}|^{2}\nu_{j\rightarrow a}^{t}}}\nonumber \\
 & \overset{a}{\approx}\frac{\sum_{a}|A_{ai}|^{2}x_{i}+A_{ai}^{*}w_{a}+A_{ai}^{*}\sum_{j\neq i}A_{aj}\bigl(x_{j}-\hat{x}_{j\rightarrow a}^{t}\bigr)}{\sum_{a}|A_{ai}|^{2}}\nonumber \\
 & \overset{b}{\approx}x_{i}+\frac{1}{M\gamma}\Bigl[\sum_{a}A_{ai}^{*}w_{a}+\sum_{a}A_{ai}^{*}\sum_{j\neq i}A_{aj}\bigl(x_{j}-\hat{x}_{j\rightarrow a}^{t}\bigr)\Bigr],\label{eq:Mean_Rit}
\end{align}
where step $\overset{a}{\approx}$ in (\ref{eq:Mean_Rit}) is due
to the assumption that $\sigma^{2}+\sum_{j\neq i}|A_{aj}|^{2}\nu_{j\rightarrow a}^{t}$
is independent of $\mu$ such that it is canceled out in the denominator
and numerator; step $\overset{b}{\approx}$ in (\ref{eq:Mean_Rit})
is attributed to the assumption that $A_{ai}$ admits i.i.d. distribution
with mean zero and variance $\gamma$ so that $\sum_{a}|A_{ai}|^{2}\approx\sum_{a}\gamma=M\gamma$. 

Denote by $r_{i}^{t}=\sum_{a}A_{ai}^{*}w_{a}+\sum_{a}A_{ai}^{*}\sum_{j\neq i}A_{aj}\bigl(x_{j}-\hat{x}_{j\rightarrow a}^{t}\bigr)$,
then $r_{i}^{t}$ is a complex random variable with respect to the
distribution of the measurement matrix elements and the complex Gaussian
noise $w_{a}\sim\mathcal{CN}\bigl(w_{a};0,\sigma^{2}\bigr)$. By central
limit theorem, it can be verified that $r_{i}^{t}$ is a complex Gaussian
random variable with zero mean and variance $M\gamma\bigl(\sigma^{2}+\gamma NE^{t}\bigr)$.
Thus, $R_{i}^{t}$ can be reformulated as
\begin{alignat}{1}
R_{i}^{t} & =x_{i}+\sqrt{\frac{\sigma^{2}+\gamma NE^{t}}{M\gamma}}z,\label{eq:rit-refor}
\end{alignat}
where $z\sim\mathcal{CN}\bigl(z;0,1\bigr)$ is a complex Gaussian
random variable with zero mean and unit variance. 

Then, from (\ref{eq:Vi_app}), we obtain that 
\begin{equation}
\Sigma_{i}^{t}\approx\frac{\sigma^{2}+\gamma NV^{t}}{M\gamma}.\label{eq:Sigma_it}
\end{equation}

Thus, the MMSE estimate of $x_{i}$ at the $\bigl(t+1\bigr)$-th iteration
is given by $f_{a}\bigl(\Sigma_{i}^{t},R_{i}^{t}\bigr)$, and the
corresponding MSE reads 
\begin{align}
E^{t+1} & =\int dx_{i}P_{0}\bigl(x_{i}\bigr)\int\mathcal{D}z\Bigl|f_{a}\bigl(\Sigma_{i}^{t},R_{i}^{t}\bigr)-x_{i}\Bigr|^{2},\label{eq:E_t_update}
\end{align}
where $P_{0}\bigl(x_{i}\bigr)$ is the prior distribution defined
in (\ref{eq:prior_dist}) and $\mathcal{D}z$ is the unit complex
Gaussian measure $\mathcal{D}z=e^{-|z|^{2}}dz/\pi.$

According to (\ref{eq:aver-var}), the average variance estimate at
the $\bigl(t+1\bigr)$-th iteration is given by 
\begin{align}
V^{t+1} & =\int dx_{i}P_{0}\bigl(x_{i}\bigr)\int\mathcal{D}zf_{c}\bigl(\Sigma_{i}^{t},R_{i}^{t}\bigr).\label{eq:V_t_update}
\end{align}

So that (\ref{eq:E_t_update}) and (\ref{eq:V_t_update}) constitute
the state evolution equations for CB-AMP.

\section{Simulation Results}

We evaluate the performance of CB-AMP for reconstruction of complex-valued
sparse signals. The elements of measurement matrix $\mathbf{A}$ are
generated using i.i.d. complex Gaussian distribution with mean zero
and variance $\gamma=1/N$. The complex-valued sparse signals are
assumed to follow $\rho$-sparse Bernoulli-Gaussian distribution,
i.e., $p_{0}\bigl(\mathbf{x}\bigr)=\stackrel[i=1]{N}{\prod}\bigl(\bigl(1-\rho\bigr)\delta\bigl(x_{i}\bigr)+\rho\mathcal{CN}\bigl(x_{i};\mu,\tau\bigr)\bigr),$
where $0<\rho<1$, $\mu$, $\tau$ are known. In this case, after
some algebra, the posterior mean and variance defined in (\ref{eq:mean_def})
and (\ref{eq:var_def}) can be calculated as
\begin{align}
f_{a}\left(R,\Sigma\right) & =\frac{m}{\frac{1-\rho}{\rho}\frac{\tau}{V}\exp\bigl(\frac{|\mu|^{2}}{\tau}-\frac{|m|^{2}}{V}\bigr)+1},\label{eq:mean_cal}\\
f_{c}\left(R,\Sigma\right) & =\frac{\rho\frac{V}{\tau}\exp\bigl(\frac{|m|^{2}}{V}-\frac{|\mu|^{2}}{\tau}-\frac{|R|^{2}}{\Sigma}\bigr)\bigl(\bigl|m\bigr|^{2}+V\bigr)}{Z\left(R,\Sigma\right)}\nonumber \\
 & \hspace{1em}-\bigl|f_{a}\left(R,\Sigma\right)\bigr|^{2},\label{eq:var_cal}
\end{align}
where 
\begin{align}
V & =\frac{\tau\Sigma}{\Sigma+\tau},\label{eq:var_V}\\
m & =\frac{\tau R+\Sigma\mu}{\Sigma+\tau},\label{eq:mean_m}\\
Z\left(R,\Sigma\right) & =\bigl(1-\rho\bigr)\exp\bigl(-\frac{|R|^{2}}{\Sigma}\bigr)\label{eq:Z_normal}\\
 & \hspace{1em}+\rho\frac{V}{\tau}\exp\bigl(\frac{|m|^{2}}{V}-\frac{|\mu|^{2}}{\tau}-\frac{|R|^{2}}{\Sigma}\bigr),
\end{align}

\begin{figure}[H]
\begin{centering}
\includegraphics[width=8cm]{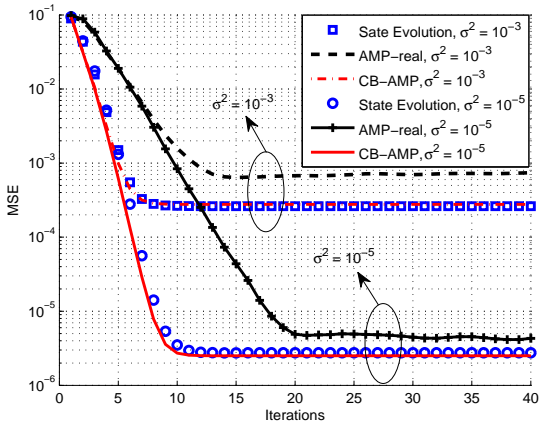}
\par\end{centering}

\protect\caption{MSE versus the number of iterations. $N=10^{3}$, $\alpha=0.5$, $\rho=0.1$,
$\mu=0,\tau=1$.}
\label{fig:MSE_performance} 
\end{figure}

In the noisy case, the MSE performances of different methods are depicted
in Fig. \ref{fig:MSE_performance} when $N=10^{3},\alpha=M/N=0.5,\rho=0.1,\mu=0,\tau=1$.
Other simulation scenarios are omitted due to lack of space. Compared
with the real AMP method, which converts the complex signal to the
real domain before processing, CB-AMP improves the MSE evidently and
converges more quickly. In addition, the theoretical state evolution
prediction matches closely with the experimental result, implying
that the performance of CB-AMP can be accurately predicted by state
evolution. 

\begin{figure}[H]
\begin{centering}
\includegraphics[width=8cm]{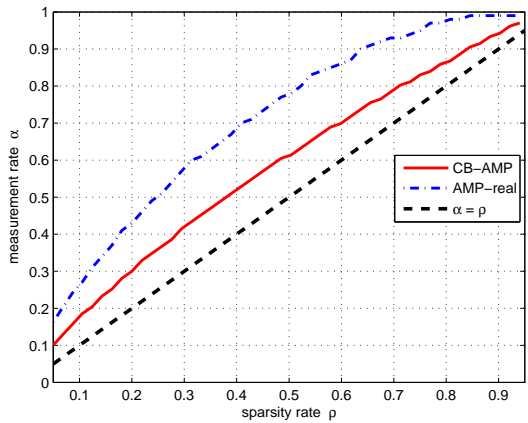}
\par\end{centering}

\protect\caption{Phase transition curve. $N=10^{3}$, $\mu=0,\tau=1$, Number of iterations
$T=500$. Success is stated if $MSE<10^{-4}$.}
\label{fig:Phase_transition} 
\end{figure}

In the noiseless case, i.e., $\sigma^{2}=0$, the phase transition
curves are shown in Fig. \ref{fig:Phase_transition}. For both real
AMP and CB-AMP, the signal length is $N=1000$, and number of iterations
is set to be $T=500$. The phase transition curves display the relationship
between the measurement rate $\alpha$ and the sparsity rate $\rho$
at a success rate of $50\%$, where the success of recovering the
original signal is stated if the mean square error $MSE<10^{-4}$.
The line $\alpha=\rho$ indicates the maximum-a-posterior (MAP) threshold.
As shown in Fig. \ref{fig:Phase_transition}, CB-AMP improves the
phase transition curve of real AMP significantly, which is attributed
to the structured sparsity of the complex signal.

\section{Conclusion}

In this paper, we considered the problem of recovering complex-valued
signals from a set of complex-valued linear measurements. Using EP,
we have extended the AMP algorithm to complex-valued Bayesian AMP
(CB-AMP). This novel perspective leads to a more concise and natural
derivation of CB-AMP, without resorting to sophisticated transformations
between the complex domain and the real domain. State evolution equations
for CB-AMP are also derived. Simulation results demonstrate that CB-AMP
outperforms real AMP in the complex-valued case and that state evolution
predicts the reconstruction performance of CB-AMP accurately.

\section*{Acknowledgments }

The first author would like to thank C. Schülke and P. Schniter for
valuable discussions on complex AMP. This work was partially supported
by the National Nature Science Foundation of China (Grant Nos. 91338101,
91438206, and 61231011), the National Basic Research Program of China
(Grant No. 2013CB329001).

\bibliographystyle{IEEEtran}
\bibliography{IEEEabrv,Mybib}

\end{document}